\documentclass{article}

\usepackage[final]{neurips_2026}
\makeatletter
\renewcommand{\@noticestring}{Workshop on Who Verifies the Agents? Toward Reliable Agent Development, NeurIPS 2026.}
\makeatother

\usepackage[utf8]{inputenc}
\usepackage[T1]{fontenc}
\usepackage{hyperref}
\usepackage{url}
\usepackage{booktabs}
\usepackage{amsfonts}
\usepackage{nicefrac}
\usepackage{microtype}
\usepackage{xcolor}
\usepackage{amsmath}
\usepackage{amssymb}
\usepackage{graphicx}
\usepackage{array}
\usepackage{tikz}
\usetikzlibrary{arrows.meta, positioning, fit, backgrounds}
\usepackage{amsthm}
\usepackage{enumitem}
\newtheorem{theorem}{Theorem}
\newtheorem{proposition}{Proposition}
\theoremstyle{definition}

\newcommand{\phivec}{\varphi}
\newcommand{\Rad}{R}

\title{An Account of Guardrail Validation via Conformal Prediction:\\What Works, What Does Not}
\workshoptitle{Who Verifies the Agents? Toward Reliable Agent Development}

\author{%
  Sherwin Varghese, Alessio Lomuscio \\
  Imperial College London\\
  \texttt{sherwin.varghese@imperial.ac.uk}\\
}

\begin{document}

\maketitle

\begin{abstract}
A guardrail classifier can be certified against a threat model measured from an
attacker one can run, using a split-conformal radius, and scored by
\emph{certified harmful coverage} (CHC): of the harmful prompts the guard already
fires on, the fraction it also certifies, at a fixed $5\%$ benign false-positive
rate. On a public toxicity guard this reaches $0.374$, and we ask what the
missing $0.626$ consists of. With an oracle difficulty estimate inside the radius,
the certified set is exactly the set of prompts no sampled edit evades, so every
fired prompt is certified, lost to the attacker, or lost to a loose difficulty
estimate; a few certified prompts ($2.4\%$ at most) are evaded anyway, which is
the conformal error budget. On two BERT-family toxicity guards the loose share is
$1.4$ to $3.4$ times the attacker's. Measuring the difficulty with $8$ of the
attacker's $32$ sampled edits raises CHC from $0.372$ to $0.455$ on the same $24$
remaining edits (paired difference $0.083$, $95\%$ CI $[0.019, 0.146]$, positive on
all four seeds), recovering about a fifth of the loose share. When the guard
changes, a certificate shared by every head in a feature space is valid and
vacuous, since its radius equals the worst head's score. A certificate for a
neighbourhood of the current head survives a change of loss and keeps little after
retraining on resampled data. Certifying the maximum of several guards needs no
union bound; it matches the depth a deployer would pick and adds nothing across
re-fits of one head.
\end{abstract}

\section{Introduction}
\label{sec:intro}

A deployed agent is gated by classifiers that screen its prompts and tool calls.
These classifiers do the verifying, and few carry a guarantee of their own. When
one does, it arrives as a single number, naturally read as a statement about the
attacker: what the certificate fails to reach is what a determined adversary
takes, and the only remedy is a more robust guard. For the guards we study, that
reading is mostly wrong.

We start from a certificate whose threat model is measured: a finite set of
strings that a runnable attacker produced from each prompt, with a split-conformal
radius bounding how far those strings move the guard's margin. In our earlier
experiments on \texttt{unitary/\allowbreak toxic-bert} it certified $44\%$ of the harmful
prompts the guard fires on, where the box-shaped pre-activation regions verified
by Kezins et al.\ \cite{kezins2026beyond} return a counterexample in every
configuration they report, as they did on this guard when we reproduced them.

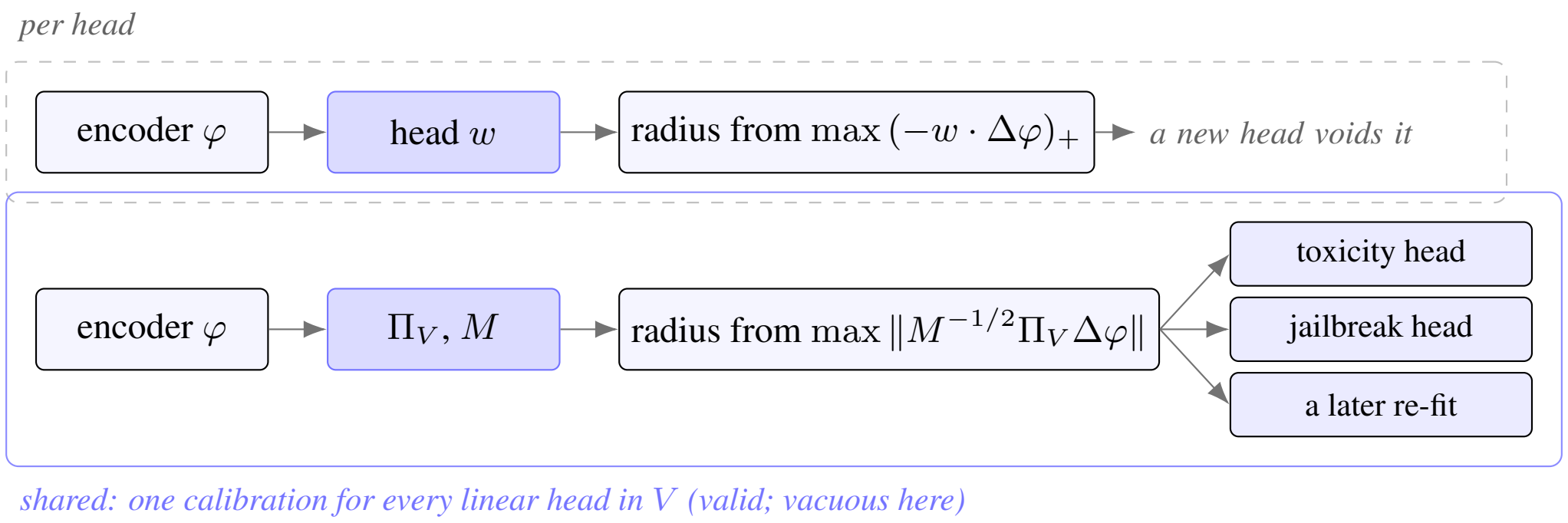
\begin{figure}[t]
\centering
\begin{tikzpicture}[
  font=\small,
  box/.style={draw, rounded corners=2pt, align=center, minimum height=7mm,
              inner xsep=3pt, fill=blue!4},
  dark/.style={box, fill=blue!14, draw=blue!55},
  head/.style={draw, rounded corners=2pt, align=center, minimum height=5.5mm,
               minimum width=26mm, inner xsep=2pt, fill=blue!8, font=\footnotesize},
  lbl/.style={align=left, font=\footnotesize\itshape, text=black!62},
  ar/.style={-{Latex[length=2mm]}, draw=black!55}]

% per head: the certificate is attached to the head
\node[box, minimum width=20mm] (p1) at (0,1.35) {encoder $\phivec$};
\node[dark, minimum width=20mm, right=5mm of p1] (p2) {head $w$};
\node[box, minimum width=32mm, right=5mm of p2] (p3)
     {radius from $\max\,(-w\cdot\Delta\phivec)_+$};
\node[lbl, right=3.5mm of p3, text width=27mm] (p4)
     {a new head voids it};
\draw[ar] (p1) -- (p2);
\draw[ar] (p2) -- (p3);
\draw[ar] (p3) -- (p4);

% shared: the certificate is attached to the representation
\node[box, minimum width=20mm] (o1) at (0,-0.35) {encoder $\phivec$};
\node[dark, minimum width=20mm, right=5mm of o1] (o2) {$\Pi_V$, $M$};
\node[box, minimum width=32mm, right=5mm of o2] (o3)
     {radius from $\max \|M^{-1/2}\Pi_V \Delta\phivec\|$};
\draw[ar] (o1) -- (o2);
\draw[ar] (o2) -- (o3);

\node[head, right=6mm of o3, yshift=6.5mm] (h1) {toxicity head};
\node[head, right=6mm of o3] (h2) {jailbreak head};
\node[head, right=6mm of o3, yshift=-6.5mm] (h3) {a later re-fit};
\draw[ar] (o3.east) -- (h1.west);
\draw[ar] (o3.east) -- (h2.west);
\draw[ar] (o3.east) -- (h3.west);

\begin{scope}[on background layer]
  \node[fit=(p1)(p4), draw=black!25, dashed, rounded corners=3pt,
        inner sep=2.5mm] (bprior) {};
  \node[fit=(o1)(o3)(h1)(h3), draw=blue!45, rounded corners=3pt,
        inner sep=2.5mm] (bours) {};
\end{scope}
\node[lbl, above=0.5mm of bprior.north west, anchor=south west] {per head};
\node[lbl, below=0.5mm of bours.south west, anchor=north west, text=blue!55]
     {shared: one calibration for every linear head in $V$ (valid; vacuous here)};
\end{tikzpicture}
\caption{Two places to attach a certificate. Scoring the worst logit drop along
$w$ ties the calibration set to one head, so a re-fit voids it. Scoring the worst
displacement inside a fixed feature space $V$ gives a radius that every linear
head in $V$ inherits on the same event, with no union bound
(Proposition~\ref{prop:shared}). The price is paid in the radius:
Proposition~\ref{prop:tight} shows it equals the worst head's score, and
\S\ref{sec:exp:portability} measures it as vacuous. The neighbourhood certificate
of \S\ref{sec:shared} sits between the two.}
\label{fig:overview}
\end{figure}

The remainder can be split exactly. The radius is $q_\alpha \hat{s}(x)$ for a
difficulty estimate $\hat{s}$, and the guarantee holds whatever $\hat{s}$ is. With
$\hat{s}$ equal to the attacker's own worst drop, $q_\alpha = 1$ and the certified
set is the set of prompts no sampled edit evades. A fired prompt that a deployed
certificate misses is then either defeated by the attacker or certifiable by a
sharper $\hat{s}$, and the few certified prompts an edit does evade are the
conformal error budget (Proposition~\ref{prop:oracle}). On our guards the loose
share is $1.4$ to $3.4$ times the attacker's, and a defender can act on it,
because $\hat{s}$ never had to be a prediction. The same accounting prices what
happens when a guard is retrained or a policy is added. Figure~\ref{fig:overview}
shows where a certificate can attach, and Table~\ref{tab:summary} summarises what
works and what does not.

\paragraph{Contributions.}
\begin{itemize}[leftmargin=*,itemsep=1pt,topsep=2pt]
\item \textbf{An exact accounting of the shortfall, and a partial repair}
(\S\ref{sec:exp:shared}). Proposition~\ref{prop:oracle} is elementary; its value
is the measurement it licenses. Measuring $\hat{s}$ with $8$ of $32$ sampled edits
raises CHC by $0.083$ $[0.019, 0.146]$ on the same certified edits, on all four
seeds and both held-out settings.
\item \textbf{Three ways of letting the guard change, each priced}
(\S\ref{sec:exp:portability}, \S\ref{sec:exp:composite}). A certificate shared by
a whole space of heads costs nothing in $\alpha$ and is vacuous, and
Proposition~\ref{prop:tight} shows its radius is the worst head's score. A
neighbourhood of the current head keeps $74\%$ of its CHC for a move of a tenth of
its norm. A maximum over guards needs no union bound
(Theorem~\ref{thm:composite}) and is paid for in the false-positive budget
(Proposition~\ref{prop:price}).
\item \textbf{Which prompts get certified, and where the geometry comes from}
(\S\ref{sec:exp:which}, \S\ref{sec:exp:pooler}). Certification tracks the
guard's starting margin, and the low-rank geometry comes from the pooler's
learned weights, which can be checked before calibration.
\end{itemize}

\section{Setup}
\label{sec:setup}

\paragraph{The guard.} A guardrail classifier is an encoder
$\phivec : \mathcal{X} \to \mathbb{R}^d$ followed by a linear decision
$w \cdot \phivec(x) + b > \theta$. Here $\phivec$ is the state the final affine map
reads: the pooled state $\tanh(W_p h_{12} + b_p)$ for BERT, and the dense layer
and $\tanh$ of RoBERTa's classification head. The bound belongs on this state,
since the classifier never scores the hidden state before it.

\paragraph{Measured threat models.} $T(x)$ is a finite set of strings, called
\emph{edits}, produced by running an attacker on $x$. The lexical attacker
samples $K = 32$ edits per prompt, each applying up to three operations of the
kind the text-attack literature uses \cite{jin2020textfooler,
morris2020textattack}: deleting, duplicating or swapping words, inserting a filler
such as ``um'', a typo, a change of case, or punctuation. The gradient attacker
runs $16$ randomised restarts of a white-box token-flip search in the manner of
HotFlip \cite{ebrahimi2018hotflip}, one edit per restart. Every edit is a string,
so no realisability gap arises; the price is that $T(x)$ samples the attacker's
own distribution, and every guarantee below inherits that.

\paragraph{Split-conformal certificates.} Let $s(x)$ be the worst drop in the
guard's margin that the attacker achieves on $x$ (\S\ref{sec:shared}). The $600$
harmful prompts are split at random into train, calibration and test folds of
$204$, $198$ and $198$. For a difficulty estimate $\hat{s}$ fixed without the
calibration fold, the radius $\Rad(x) = q_\alpha \hat{s}(x)$, with $q_\alpha$ the
$\lceil (n+1)(1-\alpha) \rceil$-th smallest value of $s/\hat{s}$ over the $n$
calibration prompts, satisfies $\mathbb{P}(s(X) \le \Rad(X)) \ge 1-\alpha$ for a
test prompt $X$ exchangeable with them \cite{papadopoulos2002inductive,
papadopoulos2008normalized, lei2018distribution}. This \emph{coverage} is marginal: it averages over the
calibration split, the test prompt and the attacker's sampling. A prompt is
\emph{certified} when its margin exceeds its radius, $h(x) > \Rad(x)$.
\emph{Certified harmful coverage} (CHC) is the fraction of harmful test prompts
the guard fires on that are certified; since it conditions on firing, we also
report the \emph{certified harmful rate} (CHR), the fraction of all harmful test
prompts that are fired on and certified.

\paragraph{Heads and thresholds.} Six linear heads are fitted on the train
harmful prompts and all $3000$ benign prompts: ridge regression on $\pm 1$
targets ($\lambda = 1$), the \emph{reference head}; ridge fits on two bootstrap
resamples ($\lambda = 1$ and $10$); ridge with harmful prompts weighted threefold;
a hinge-loss fit started from the ridge solution; and ridge on ToxicChat's
jailbreaking flag, which marks $161$ of our $600$ toxic prompts and no benign one.
Each threshold $\theta$ is the $95$th percentile of the head's scores on the same
benign prompts, so the $5\%$ false-positive rate is in sample; refitting on half
of them gives $5.1\%$ on the other half (Table~\ref{tab:floor}). CHC is the mean
over the six heads unless stated. The \emph{shipped head}, toxic-bert's own
weight vector, serves only as a diagnostic.

\paragraph{The displacement subspace.} Displacements
$\Delta = \phivec(x') - \phivec(x)$ are sharply low rank: on toxic-bert, $7$ of
$768$ directions carry $90\%$ of the lexical attacker's displacement energy
(Table~\ref{tab:pooler}). In the standardised coordinates the certificate uses,
the leading $k=16$ train directions $U_k$ carry $90\%$; with
$P = U_k U_k^{\!\top}$ and $Q = I - P$, every head is placed inside
$\mathrm{range}(Q)$. In our earlier experiments on the same guard this raised CHC
from $0.253$ for the shipped head to $0.440$ at no cost in accuracy. We take the
construction as given and re-measure its geometry (\S\ref{sec:exp:pooler}).

\section{What a certificate pays for}
\label{sec:shared}

Write a guard's normalised margin as
$h(x) = (w \cdot \phivec(x) + b - \theta)/\sigma$, with $\sigma$ the standard
deviation of the head's benign scores, so the guard fires exactly when $h(x) > 0$.
Let $s(x) = \max_{x' \in T(x)} (h(x) - h(x'))_+$ be the worst drop the attacker
achieves. Both $\theta$ and $\sigma$ are fixed before calibration, which makes them
part of the function being certified. The estimate $\hat{s}$ is floored at $0.05$
in raw logit units to keep the ratio finite. Apart from that floor, rescaling by
$\sigma$ changes neither $s/\hat{s}$ nor which prompts are certified, so per-head
certificates are computed on raw logits, and $\sigma$ enters only when guards are
composed (\S\ref{sec:composition}).

\subsection{The accounting}

Validity survives any choice of $\hat{s}$; sharpness does not. The next
proposition measures how much sharpness is at stake.

\begin{proposition}[the oracle radius and the accounting; proof in Appendix~\ref{app:proofs}]
\label{prop:oracle}
Take $\hat{s} = s$. Then $q_\alpha = 1$ and $\Rad(x) = s(x)$, and a fired prompt
is certified if and only if no edit in $T(x)$ evades the guard; the fraction of
fired prompts with this property is the \emph{empirical ceiling}. For any
$\hat{s}$, let $E$ be the event that no edit evades and $C$ the event that the
prompt is certified, both among fired test prompts. Then
\begin{equation}
1 - \mathrm{CHC}
= \underbrace{\mathbb{P}(\bar{E} \cap \bar{C})}_{\text{attacker}}
+ \underbrace{\mathbb{P}(E \cap \bar{C})}_{\text{loose}} ,
\qquad
\text{ceiling} - \mathrm{CHC}
= \mathbb{P}(E \cap \bar{C}) - \mathbb{P}(\bar{E} \cap C) ,
\label{eq:accounting}
\end{equation}
and every prompt in $\bar{E} \cap C$ has $s(x) > \Rad(x)$, so this last group is
part of the miscoverage the conformal guarantee allows.
\end{proposition}

The proposition is elementary once $\hat{s} = s$ is chosen; its use is that it
splits one number into parts a defender can act on. Prompts in the loose share
are evaded by no sampled edit, and only the difficulty estimate stands between
them and a certificate. Prompts in $\bar{E} \cap C$ are certified although an edit
evades them; a two-way split that reads $1 - \text{ceiling}$ as the attacker's
share books them to the attacker, and so understates the loose share by exactly
that group. The oracle reads the test prompt's own score and is no certificate,
so the loose share bounds, up to the error budget, what a better $\hat{s}$ can
recover.

Since $\hat{s}$ need not be a prediction, a defender can measure it by running
the attacker a few times. Split the $K$ sampled edits of each prompt into $K_0$
that feed $\hat{s}$ and $K - K_0$ that feed $s$; the ratio stays exchangeable
across prompts, so the certificate stays valid. The certified threat model
shrinks to $K - K_0$ edits, and deployment runs the attacker $K_0$ times on every
incoming prompt.

\subsection{Quantifying over a space of heads}

Fix an $m$-dimensional subspace $V \subseteq \mathbb{R}^d$ with orthogonal
projector $\Pi_V$, and a positive definite $M$ on $V$, both before calibration.
Score each prompt by the worst displacement inside $V$, measured in the inner
product $M$ induces:
\begin{equation}
s_{V,M}(x) \;=\; \max_{x' \in T(x)} \big\| M^{-1/2} \Pi_V (\phivec(x') - \phivec(x)) \big\|_2 .
\label{eq:score}
\end{equation}

\begin{proposition}[shared certificate; proof in Appendix~\ref{app:proofs}]
\label{prop:shared}
With probability at least $1-\alpha$ over the test prompt $X$, simultaneously
for every $w \in V$ and every $x' \in T(X)$,
$w \cdot \phivec(x') \ge w \cdot \phivec(X) - \Rad_{V,M}(X)\,\|M^{1/2} w\|_2$.
The heads may be chosen after the calibration set is drawn.
\end{proposition}

The random event $\{s_{V,M}(X) \le \Rad_{V,M}(X)\}$ concerns the prompt alone,
and on it the inequality holds for every $w \in V$ at once, so certifying many
heads costs nothing in $\alpha$. The cost appears in the radius, and no sharper
argument reduces it.

\begin{proposition}[the price is intrinsic; proof in Appendix~\ref{app:proofs}]
\label{prop:tight}
For $w \in V$ with $\|M^{1/2}w\|_2 = 1$, let
$s_w(x) = \max_{x' \in T(x)} (-\Delta \cdot w)_+$ be the worst logit drop of head
$w$. For each edit, $\|M^{-1/2}\Pi_V \Delta\|_2 =
\sup \{ -\Delta \cdot w : w \in V,\ \|M^{1/2}w\|_2 \le 1 \}$, attained at
$w \propto -M^{-1}\Pi_V \Delta$. Consequently $s_{V,M}(x) = \sup_{w} s_w(x)$ for
every prompt, and a radius valid for all of $V$ on one event can be no smaller
than the shared score.
\end{proposition}

Covering a space of heads means covering, prompt by prompt, the head the
attacker hurts most. In the experiments $V$ is spanned by the leading generalised
eigenvectors of $\Sigma_R v = \lambda \Sigma_D v$ inside $\mathrm{range}(Q)$, with
$\Sigma_R$ the covariance of clean representations and $\Sigma_D$ that of
displacements on train, and $M = \Sigma_D$ restricted to $V$, so each head is
charged by the spread of its own logit under the attacker.

\subsection{Quantifying over a neighbourhood}

Retraining moves a head a limited distance, so it is enough to cover
$\mathcal{W}_\rho = \{w_{\mathrm{ref}} + \delta : \delta \in V,\ \|M^{1/2}\delta\|_2 \le \rho\}$,
and we report $\rho$ as a fraction of $\|M^{1/2} w_{\mathrm{ref}}\|_2$.

\begin{proposition}[neighbourhood certificate; proof in Appendix~\ref{app:proofs}]
\label{prop:ball}
Let $s_\rho(x) = \max_{x' \in T(x)} \big[ (-\Delta \cdot w_{\mathrm{ref}})_+ + \rho \|M^{-1/2}\Pi_V \Delta\|_2 \big]$
and let $\Rad_\rho$ be its split-conformal radius. Then with probability at
least $1-\alpha$, every $w \in \mathcal{W}_\rho$ satisfies
$-\Delta \cdot w \le \Rad_\rho(X)$ for every $x' \in T(X)$. At $\rho = 0$ this
is the per-head certificate.
\end{proposition}

The added charge is linear in $\rho$, scaled by the shared score that
Proposition~\ref{prop:tight} prices.

\subsection{Composition}
\label{sec:composition}

The third option runs the old guard next to the new one, a shape agent stacks
often have already. Given members $h_1, \dots, h_J$, each a normalised margin,
let
\begin{equation}
h(x) \;=\; \max_{j \le J} h_j(x) ,
\label{eq:composite}
\end{equation}
which fires exactly when some member fires. The member that certifies a prompt
is chosen after the prompt is seen, which suggests a union bound. None is needed.

\begin{theorem}[composition is free in $\alpha$; proof in Appendix~\ref{app:proofs}]
\label{thm:composite}
Let $s_h$ be the worst drop of $h$ and $\Rad_h$ its split-conformal radius at
level $\alpha$. Then
\begin{enumerate}[leftmargin=*,itemsep=0pt,topsep=2pt]
\item \textbf{Validity.} $h$ is a fixed function of the input, through the frozen
encoder, so $\mathbb{P}(s_h(X) \le \Rad_h(X)) \ge 1-\alpha$ with no dependence on
$J$ and no correction for the choice of member.
\item \textbf{Dominance.} For every prompt,
$s_h(x) \le \min_{j} \big[ s_j(x) + (h(x) - h_j(x)) \big]$, and therefore
\begin{equation}
h(x) - s_h(x) \;\ge\; \max_{j \le J} \big[ h_j(x) - s_j(x) \big] .
\label{eq:dominance}
\end{equation}
\end{enumerate}
\end{theorem}

Inequality~\eqref{eq:dominance} bounds the composite's worst sampled margin. The
certificate also depends on the radius, calibrated on $s_h$, and on member
thresholds that tighten. When members disagree about which prompts are hard, the
composite's worst-case margin beats that of any fixed member.

\begin{proposition}[the price of composition; proof in Appendix~\ref{app:proofs}]
\label{prop:price}
Let each member fire on a fraction $\beta$ of benign prompts. The composite's
benign rate lies in $[\beta, \min(1, J\beta)]$, attaining the lower end when the
members' benign firing sets coincide and the upper end when they are disjoint.
Holding the composite at a target rate $a$ therefore forces $\beta \le a$, and
\eqref{eq:dominance} holds with each $h_j$ evaluated at the tightened threshold.
\end{proposition}

A useful member should therefore fire on the same benign prompts as the others,
so that it costs little threshold, and be defeated by different edits, so that it
is the maximum where the others lose. Both are correlations measurable before
calibration.

\begin{table}[t]
\centering
\caption{What each construction certifies, what else voids it besides a change of
encoder or attacker, its cost per prompt at deployment, and the verdict on
toxic-bert with the lexical attacker. Every guarantee holds with probability at
least $1-\alpha$, marginally, over the attacker's sampled edits.}
\label{tab:summary}
\footnotesize
\setlength{\tabcolsep}{3pt}
\begin{tabular}{>{\raggedright\arraybackslash}p{2.35cm}>{\raggedright\arraybackslash}p{2.9cm}>{\raggedright\arraybackslash}p{2.0cm}>{\raggedright\arraybackslash}p{1.8cm}>{\raggedright\arraybackslash}p{3.6cm}}
\toprule
construction & certifies & also voided by & cost & verdict here \\
\midrule
per head, predicted $\hat{s}$ & one head & a new head & none & CHC $0.374$ \\
per head, measured $\hat{s}$ & one head, on $K - K_0$ edits & a new head & $K_0$ attacker runs & CHC $0.455$, the best \\
shared over $V$ & every head in $V$ & a change of $V$, $M$ & none & valid but vacuous (CHC $0.004$) \\
neighbourhood $\mathcal{W}_\rho$ & heads within $\rho$ of the reference & a move beyond $\rho$ & as its $\hat{s}$ & survives a loss change; $\le 25\%$ left after resampling \\
maximum of $J$ guards & the composite & any member & $J$ heads & matches train-picked depth; no gain over re-fits \\
\bottomrule
\end{tabular}
\end{table}

\section{Experiments}
\label{sec:exp}

We use two frozen public guards from one publisher \cite{detoxify}, built on
BERT and RoBERTa \cite{devlin2019bert, liu2019roberta}:
\texttt{unitary/\allowbreak toxic-bert} and
\texttt{unitary/\allowbreak unbiased-\allowbreak toxic-\allowbreak roberta}. Data
are $600$ harmful and $3000$ benign ToxicChat prompts
\cite{lin2023toxicchat}, set up as in \S\ref{sec:setup}, at $5\%$ benign
false-positive rate, $\alpha = 0.05$ and $m = 64$ unless stated. 
% Of thirteen
% predictions registered before the runs, eight were falsified as stated and two
% held in part (Appendix~\ref{app:prereg}). 
Measured $\hat{s}$ and the depth
composites were explored on toxic-bert with the lexical attacker at seed $0$, with
the gradient attacker and the second guard held out. Intervals, decompositions and
sensitivity checks were added after review and were not registered
(Appendix~\ref{app:review}).

\begin{figure}[t]
\centering
\includegraphics[width=0.52\textwidth]{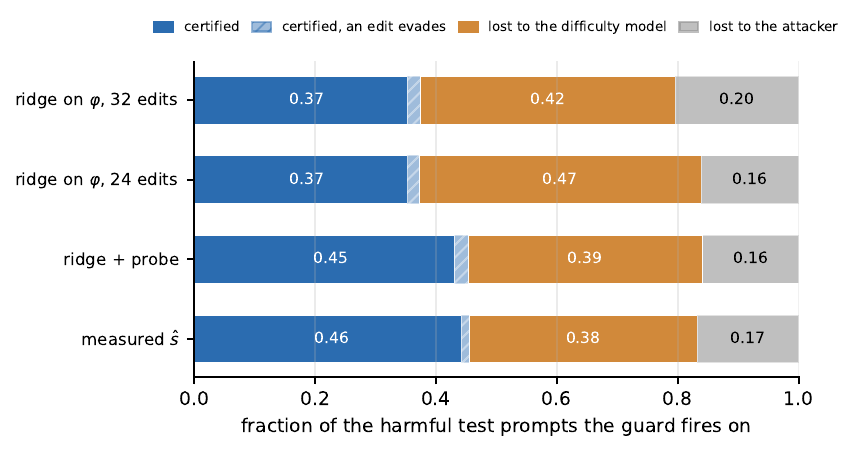}
\hfill
\includegraphics[width=0.42\textwidth]{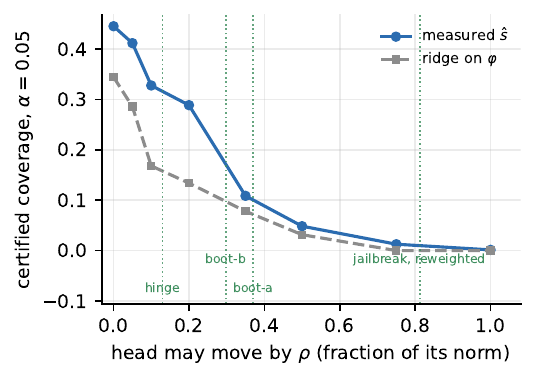}
\caption{Left: where the harmful prompts a guard fires on go, as the difficulty
estimate changes (toxic-bert, lexical attacker, mean over six heads). The orange
band is the loose share of Proposition~\ref{prop:oracle}; it is the larger loss in
every row. The hatched sliver is certified although an edit evades it. Right: the
CHC a certificate keeps as the head is allowed to move, with the distances real
re-fits move marked. Measuring $\hat{s}$ lifts the whole curve.}
\label{fig:accounting}
\end{figure}

\subsection{The accounting, and measured difficulty}
\label{sec:exp:shared}

Table~\ref{tab:difficulty} and Figure~\ref{fig:accounting} measure each part of
Proposition~\ref{prop:oracle}; only $\hat{s}$ changes between rows. The baseline,
a ridge regression of $\log s$ on $\phivec$, certifies $0.374$ $[0.31, 0.44]$
against an empirical ceiling of $0.775$. Of the $0.626$ it misses, the attacker
accounts for $0.204$ and the loose estimate for $0.422$, a ratio of $2.1$; another
$0.021$ is certified although an edit evades it. Across seeds, attackers and
guards the ratio runs from $1.4$ to $3.4$ (Table~\ref{tab:accounting}). The oracle
certifies $0.798$ against a ceiling of $0.819$; the floor under $\hat{s}$ causes the
gap, mostly on the jailbreak head, whose logits are on a smaller scale ($0.083$
there, at most $0.017$ elsewhere).

\paragraph{Measuring the difficulty.} Measuring the worst drop on $K_0 = 8$ of the
$32$ edits and certifying the other $24$ gives $0.455$, against $0.372$ for a ridge
estimate on the same $24$: a paired difference of $0.083$ $[0.019, 0.146]$
(Table~\ref{tab:paired}). The smaller threat model does not explain it, since the
ridge certifies $0.374$ on all $32$ edits. The gain recovers $0.090$ of the $0.467$
loose share, about a fifth (a ninth to a third across seeds). The neighbourhood
certificate gains more, from $0.168$ to $0.328$ at $\rho = 0.10$. With $K_0 = 4$
and $16$ the gain is $0.049$ and $0.082$ (Table~\ref{tab:measure}). Fed to the
ridge as one more column, the measurement gives the same per-head CHC ($0.453$)
and only $0.193$ for the neighbourhood certificate.

\paragraph{Held-out settings and seeds.} The per-head gain transfers to the
gradient attacker, $+0.060$ with interval $[-0.001, 0.124]$, which touches zero,
and to the RoBERTa guard, $+0.104$ with interval $[0.045, 0.163]$; Holm-adjusted
one-sided $p$-values across the three settings are $0.016$, $0.027$ and $0.001$.
The registered neighbourhood gain of $0.05$ held on the gradient attacker
($+0.142$) and failed on the RoBERTa guard ($+0.020$, interval $[-0.020, 0.064]$). Over four seeds the per-head
difference is $+0.083$, $+0.092$, $+0.052$ and $+0.174$, positive each time, though
four seeds cannot give a sign test below $p = 0.0625$. Per-head coverage averages
$0.956$ over seeds, and single seeds stay inside the $[0.91, 0.99]$ that one
calibration split allows (Appendix~\ref{app:coverage}).

\subsection{How far a certificate travels when the head changes}
\label{sec:exp:portability}

\paragraph{Quantifying over a head space is vacuous.} With one calibration shared
by every head in $V$, CHC is $0.004$ at $m=64$ and $0.000$ from $m=128$ on, where
the same heads certified one at a time average $0.455$ and $0.463$
(Table~\ref{tab:shared}). The generalised eigenbasis cuts the displacement energy
inside $V$ from $0.781$ (principal components) to $0.028$ at $m=128$ and raises
per-head CHC, and the shared CHC stays at zero, as Proposition~\ref{prop:tight}
predicts: the shared score is the worst head's score.

\paragraph{A neighbourhood survives small moves.} At $\rho = 0$ the curve of
Table~\ref{tab:ball} reproduces the per-head certificate. A move of $5\%$ of the
reference head's norm keeps $93\%$ of its CHC ($0.412$ against $0.445$), $10\%$
keeps $74\%$ and $20\%$ keeps $65\%$. With the ridge estimate of $\hat{s}$, $\rho =
0.10$ keeps $49\%$, so this certificate depends on measuring the difficulty.

\paragraph{What a real re-fit costs.} A hinge-loss re-fit moves the head by $0.13$
of its norm and is first covered at $\rho = 0.20$, keeping $65\%$ of the CHC.
Bootstrap re-fits move it $0.30$ and $0.37$ and are covered at $\rho = 0.35$ and
$0.50$, keeping $25\%$ and $11\%$. The reweighted and jailbreak heads sit $0.81$
away, where nothing is left.

\subsection{Composing guards}
\label{sec:exp:composite}

Every arm in Table~\ref{tab:composite} is held to the same $5\%$ benign rate, and
the baseline is the member a deployer would pick on the train split.

\paragraph{Across depths, composition matches selection.} On the explored
setting, layers $\{1,2,3,4\}$ certify $0.700$ against $0.635$ for the train-picked
depth. That margin came from choosing among seven member sets and did not
transfer: the same composite gives $0.632$ against $0.637$ on the gradient attacker
and $0.636$ against $0.648$ on the second guard, short of the registered $0.03$.
The composite lands within $0.02$ of the depth a deployer would pick without
anyone picking one, and needs the encoder only up to layer $4$; its member set,
though, was itself chosen on one setting.

\paragraph{Across versions of a head, composition adds nothing.} These composites
normalise members by $\sigma$, which applies the floor under $\hat{s}$ in
normalised units, so the reference head alone certifies $0.353$ here against
$0.445$ in Table~\ref{tab:ball} (Table~\ref{tab:floor}). Composing it with a re-fit
never beats the better member: $0.367$ against $0.383$ with the hinge re-fit, and
$0.342$ against $0.357$ with a bootstrap re-fit. With the jailbreak head it gives
$0.407$, below that head's $0.597$, which is measured over fewer fired prompts.

\paragraph{The two failures have opposite causes.} Neither natural member set
satisfies Proposition~\ref{prop:price}'s two conditions. Depths have different
displacement subspaces (mean principal-angle cosine $0.12$ between layers $1$ and
$12$) and fire on largely different benign prompts, so holding the union at $5\%$
cuts each member to $2.4\%$ for $\{1,2,3,4\}$ and $1.5\%$ for all eight. Re-fits of
one head agree on benign scores ($0.92$ to $0.98$; adding one costs $5.0\%$ to
$4.7\%$) and on drops ($0.89$ to $0.99$, median $0.94$), so they fail together.
Only the jailbreak head decorrelates on both, at about $0.66$ and $0.64$, and only
it improves on its partner. Its flag marks a subset of ToxicChat's toxic prompts,
so it enforces a narrower policy nested in the first. Members worth composing
differ by construction, and both correlations can be computed before calibration.

\begin{table}[t]
\centering
\caption{Where the fired harmful prompts go as $\hat{s}$ changes (toxic-bert, lexical, $m=64$, mean over six heads). \emph{edits}: how many of the $32$ the certificate covers. \emph{attacker}: an edit evades, not certified; \emph{loose}: no edit evades, not certified; \emph{evaded}: certified although an edit evades. CHC, attacker and loose sum to one. Brackets: $95\%$ bootstrap intervals over test prompts. \emph{coverage}: per-head $\mathbb{P}(s \le R)$ on the $198$ harmful test prompts, seed $0$ (four-seed mean). The oracle is not a certificate.}
\label{tab:difficulty}
\footnotesize
\setlength{\tabcolsep}{3.5pt}
\begin{tabular}{lrrrrrrrr}
\toprule
$\hat{s}$ & edits & CHC & attacker & loose & evaded & ceiling & $\rho{=}0.10$ & coverage \\
\midrule
ridge on $\varphi$ & 32 & 0.374 {\scriptsize[0.31, 0.44]} & 0.204 & 0.422 & 0.021 & 0.775 & 0.176 & 0.933 (0.951) \\
ridge on $\varphi$ & 24 & 0.372 {\scriptsize[0.31, 0.44]} & 0.161 & 0.467 & 0.020 & 0.819 & 0.168 & 0.929 (0.949) \\
ridge, probe as a feature & 24 & 0.453 {\scriptsize[0.39, 0.52]} & 0.159 & 0.388 & 0.022 & 0.819 & 0.193 & 0.934 (0.943) \\
probe alone (measured) & 24 & 0.455 {\scriptsize[0.39, 0.52]} & 0.168 & 0.377 & 0.014 & 0.819 & 0.328 & 0.943 (0.956) \\
oracle $\hat{s}=s$ & 24 & 0.798 {\scriptsize[0.74, 0.85]} & 0.181 & 0.021 & 0.000 & 0.819 & 0.655 & 1.000 (1.000) \\
\bottomrule
\end{tabular}
\end{table}

\begin{table}[t]
\centering
\caption{Certifying a neighbourhood of the reference head ($m=64$, measured $\hat{s}$). $\rho$: how far the head may move, as a fraction of its $M$-norm. CHC: mean over the heads inside the ball (last column: r ridge, h hinge, a/b bootstrap, w reweighted, j jailbreak), with a $95\%$ bootstrap interval. \emph{coverage}: of the ball's score, seed $0$ (four-seed mean); one split alone ranges over $[0.91, 0.99]$ (Appendix~\ref{app:coverage}).}
\label{tab:ball}
\small
\begin{tabular}{rrrrl}
\toprule
$\rho$ & CHC & worst & coverage & heads \\
\midrule
0.00 & 0.445 {\scriptsize[0.36, 0.53]} & 0.445 & 0.944 (0.960) & \texttt{r} \\
0.05 & 0.412 {\scriptsize[0.33, 0.50]} & 0.412 & 0.934 (0.953) & \texttt{r} \\
0.10 & 0.328 {\scriptsize[0.25, 0.41]} & 0.328 & 0.934 (0.955) & \texttt{r} \\
0.20 & 0.289 {\scriptsize[0.21, 0.37]} & 0.286 & 0.924 (0.941) & \texttt{r,h} \\
0.35 & 0.109 {\scriptsize[0.06, 0.17]} & 0.100 & 0.934 (0.951) & \texttt{r,b,h} \\
0.50 & 0.048 {\scriptsize[0.02, 0.09]} & 0.025 & 0.934 (0.944) & \texttt{r,a,b,h} \\
0.75 & 0.013 {\scriptsize[0.00, 0.03]} & 0.008 & 0.934 (0.937) & \texttt{r,a,b,h} \\
1.00 & 0.001 {\scriptsize[0.00, 0.00]} & 0.000 & 0.944 (0.936) & \texttt{r,a,b,w,h,j} \\
\midrule
\multicolumn{5}{l}{\footnotesize how far a real re-fit moves: h 0.13, b 0.30, a 0.37, j 0.81, w 0.81} \\
\bottomrule
\end{tabular}
\end{table}

\begin{table}[t]
\centering
\caption{Composing depths (toxic-bert, lexical attacker, the setting the member sets were explored on), every arm held to the same $5\%$ benign rate. \emph{cost} is the fraction of the encoder the members need between them. \emph{per-member} is the rate each member is allowed so the union meets the target, which is the price Proposition~\ref{prop:price} charges. CHR, the certified harmful rate, is the fraction of all harmful test prompts that are fired on and certified; unlike CHC it does not depend on which prompts a guard fires on. The last column is the member of that set a deployer would have picked using train data alone.}
\label{tab:composite}
\small
\begin{tabular}{lrrrrrr}
\toprule
members & cost & per-member & ceiling & CHC & CHR & train-picked depth \\
\midrule
$\{4\}$ & 0.39 & 5.0\% & 0.797 & 0.654 & 0.439 & 0.654 \\
$\{1,4\}$ & 0.39 & 3.1\% & 0.779 & 0.649 & 0.429 & 0.654 \\
$\{1,3,4\}$ & 0.39 & 2.7\% & 0.754 & 0.669 & 0.439 & 0.635 \\
$\{1,2,3,4\}$ & 0.39 & 2.4\% & 0.746 & \textbf{0.700} & 0.460 & 0.635 \\
$\{4,12\}$ & 1.00 & 3.0\% & 0.770 & 0.601 & 0.449 & 0.662 \\
$\{1,12\}$ & 1.00 & 3.0\% & 0.745 & 0.453 & 0.313 & 0.423 \\
$\{1,2,3,4,6,8,10,12\}$ & 1.00 & 1.5\% & 0.774 & 0.596 & 0.439 & 0.635 \\
\bottomrule
\end{tabular}
\end{table}

\subsection{Which prompts a certificate reaches}
\label{sec:exp:which}

For the reference head with the ridge-plus-probe estimate (third row of
Table~\ref{tab:difficulty}, seed $0$, raw logit units), $59$ of the $119$ fired
test prompts are certified. Certified and uncertified prompts suffer similar
worst drops (quartiles $0.04/0.07/0.17$ against $0.04/0.06/0.13$) and receive
similar radii ($0.16/0.22/0.25$ against $0.17/0.20/0.33$). The starting margin
separates them: $0.27/0.34/0.52$ against $0.07/0.09/0.15$. Split at the median
radius and margin, large-margin prompts are certified at $95\%$ (small radius)
and $82\%$ (large radius), small-margin prompts at $18\%$ and $0\%$. Certification
mostly fails where the guard was barely firing. Raising margins at a fixed
false-positive rate is a second route to more CHC, though adversarial training,
the obvious way to raise robust margins, lowered it here (Appendix~\ref{app:extra}).

\subsection{Where the geometry comes from}
\label{sec:exp:pooler}

The constructions need displacements at $\phivec$ to be low rank, measured by
rank$_{90}$, the number of directions carrying $90\%$ of the displacement energy:
$72$ to $214$ in the encoder's hidden states and $7$ at $\phivec$
(Table~\ref{tab:pooler}). The pooler's learned linear map does the compressing,
taking rank$_{90}$ from $74$ to $3$ at the last layer before its $\tanh$ brings it
to $7$; a random projection of matched norm followed by the same $\tanh$ leaves
$63$. This can be checked before calibration, for guards that have such a pooler.
Reading earlier layers also certifies more: layers $0$ to $2$ give $0.723$ $[0.642,
0.796]$ for $23\%$ of the encoder, against $0.541$ for the final layer
(Appendix~\ref{app:depth}).

\section{Related work}
\label{sec:related}

\paragraph{Guardrails.} Deployed language-model systems gate inputs, tool calls
and outputs with classifiers \cite{inan2023llamaguard, rebedea2023nemo, detoxify,
markov2023moderation}, evaluated on real traffic \cite{lin2023toxicchat} and
attacked with hand-written jailbreaks and gradient search over adversarial
suffixes \cite{zou2023gcg, chao2023pair}. Erase-and-check
\cite{kumar2023erasecheck} and randomised smoothing over deletions
\cite{robey2023smoothllm} certify against an edit budget at a cost on every query.
A certificate with predicted $\hat{s}$ is priced once at calibration; with
measured $\hat{s}$ it pays $K_0$ attacker runs per query, a cost of the same kind.

\paragraph{Certified robustness for text.} Interval bound propagation
\cite{gowal2018ibp}, applied to word and symbol substitutions
\cite{jia2019certified, huang2019achieving}, linear-relaxation bounds
\cite{zhang2018crown, xu2020autolirpa}, convex relaxations \cite{wong2018provable,
salman2019convex} and randomised smoothing \cite{cohen2019smoothing,
ye2020safer, zeng2023certified} certify against a ball or substitution set in an
input or embedding space. Kezins et al.\ \cite{kezins2026beyond} verify convex
regions enclosing harmful prompts in a guard's pre-activation space; their boxes
return a counterexample in every configuration, and their Gaussian mixtures give
probabilistic coverage. Such regions contain points no string maps to, the gap a
measured threat model closes; exact SMT or mixed-integer procedures
\cite{katz2017reluplex, tjeng2019mip} share it. A measured threat model is close
to statistical robustness estimation \cite{webb2019statistical}, which estimates
how often inputs drawn from a given distribution violate a property; the
conformal step turns such measurements into a radius with finite-sample coverage
on fresh prompts. For norm balls, Yang et al.\ \cite{yang2022ensemble} show that
an ensemble certifies more when its members' gradients are diverse and their
margins large. Theorem~\ref{thm:composite} is the counterpart for a maximum of
guards, where diversity means disagreement about which prompts are hard.

\paragraph{Conformal prediction.} Split conformal gives distribution-free,
finite-sample, marginal coverage for a fixed predictor
\cite{papadopoulos2002inductive, vovk2005algorithmic, lei2018distribution},
with normalised scores \cite{papadopoulos2008normalized, lei2018distribution},
group-conditional variants \cite{vovk2012mondrian} and known limits on
conditional coverage \cite{barber2021limits}. Gendler et al.\
\cite{gendler2022rscp} keep coverage when the test input itself is perturbed
inside an $\ell_2$ ball, by smoothing the score. Our test prompts are clean and
the attacker sits inside the score $s$, so exchangeability holds unchanged. Reuse of a calibration set is studied under shift in the
data \cite{tibshirani2019covariate}; when the model changes, the usual answer is
to recalibrate. Here the model moves, a head-independent score absorbs the
multiplicity, and Proposition~\ref{prop:tight} prices it. Probing and early-exit
work \cite{alain2017probing, tenney2019bert, xin2020deebert} asks which depth
separates best; Appendix~\ref{app:depth} asks which certifies best.

\section{Conclusion}
\label{sec:conc}

A certificate that reaches $37\%$ of the prompts a guard fires on looks like a
statement about the attacker. For the guards measured here, most of the shortfall
belongs to the difficulty estimate: the loose share is $1.4$ to $3.4$ times the
attacker's. Measuring the estimate recovers about a fifth of it for $K_0$ attacker
runs per prompt, and makes a certificate for a neighbourhood of heads usable. A
certificate over a whole space of heads is vacuous, because its radius is the
worst head's score; a maximum over guards costs nothing in $\alpha$ and is paid
for in the false-positive budget. A better $\hat{s}$ is the next thing to work on,
then nonlinear heads.

\paragraph{For someone shipping a guarded agent.} Quote a certificate with its
attacker and number of edits. If a few attacker runs per prompt are affordable,
measure the difficulty; here it is the largest single gain. Expect a certificate
to survive a change of loss; a refit on resampled data moves the head two to three
times as far and keeps a quarter of the CHC at best. When the guard changes,
certifying the maximum of old and new costs nothing in $\alpha$ and adds nothing
for re-fits. Before building a shared or neighbourhood certificate, check that the
pooler compresses the displacements. Under these certificates the head may change
within a declared space or radius; the encoder and the attacker may not change
without recalibration.

\section*{Limitations}
\label{sec:lim}

\paragraph{The specification is an attacker.} Coverage holds over the attacker's
sampling distribution, which reaches only the strings it tends to produce, so a
certificate quoted without its attacker is meaningless, and a new attacker needs
a new calibration set. Measuring $\hat{s}$ costs $8$ attacker runs per incoming
prompt: eight forward passes for the lexical attacker, and eight restarts of
forward and backward passes for the gradient one.

\paragraph{Linear heads, frozen encoder.} Propositions~\ref{prop:shared}
and~\ref{prop:ball} cover linear heads in $V$ only. $U_k$, $V$ and the radius are
properties of $\phivec$, and fine-tuning the encoder invalidates them; a small
update can rotate the displacement subspace without moving accuracy, so no cheap
test detects it.

\paragraph{Two guards from one publisher, one corpus, English only.} Both guards
are BERT-family Detoxify classifiers on ToxicChat. Decoder-only guards such as
Llama Guard \cite{inan2023llamaguard} and mean-pooled guards have no pooler, so
the rank$_{90}$ test is undefined for them, and we have not measured whether the
accounting carries over. Our second policy is nested in the first, while a real
stack enforces independent ones. The lexical attacker writes English with
Latin-script typos, so its geometry says nothing about other scripts.

\paragraph{In-sample thresholds, a tuned floor, one split per seed.} Thresholds
are fitted on the benign prompts they are evaluated on, though a held-out half
gives $5.1\%$. The floor under $\hat{s}$ moves CHC by up to $0.06$, and the $0.05$
used throughout is the best of the three values tried (Table~\ref{tab:floor}).
Coverage is guaranteed on average over splits; on seed $3$ the shared score misses
$27$ of $198$ test prompts where $9$ are expected (probability $6 \times 10^{-4}$),
and we found no cause (Appendix~\ref{app:coverage}).

\bibliographystyle{plain}
\bibliography{references}

@inproceedings{wong2018provable,
  title     = {Provable Defenses against Adversarial Examples via the Convex Outer Adversarial Polytope},
  author    = {Wong, Eric and Kolter, J. Zico},
  booktitle = {International Conference on Machine Learning (ICML)},
  year      = {2018}
}

@article{gowal2018ibp,
  title   = {On the Effectiveness of Interval Bound Propagation for Training Verifiably Robust Models},
  author  = {Gowal, Sven and Dvijotham, Krishnamurthy and Stanforth, Robert and Bunel, Rudy and Qin, Chongli and Uesato, Jonathan and Arandjelovic, Relja and Mann, Timothy and Kohli, Pushmeet},
  journal = {arXiv preprint arXiv:1810.12715},
  year    = {2018}
}

@inproceedings{zhang2018crown,
  title     = {Efficient Neural Network Robustness Certification with General Activation Functions},
  author    = {Zhang, Huan and Weng, Tsui-Wei and Chen, Pin-Yu and Hsieh, Cho-Jui and Daniel, Luca},
  booktitle = {Advances in Neural Information Processing Systems (NeurIPS)},
  year      = {2018}
}

@inproceedings{xu2020autolirpa,
  title     = {Automatic Perturbation Analysis for Scalable Certified Robustness and Beyond},
  author    = {Xu, Kaidi and Shi, Zhouxing and Zhang, Huan and Wang, Yihan and Chang, Kai-Wei and Huang, Minlie and Kailkhura, Bhavya and Lin, Xue and Hsieh, Cho-Jui},
  booktitle = {Advances in Neural Information Processing Systems (NeurIPS)},
  year      = {2020}
}

@inproceedings{katz2017reluplex,
  title     = {Reluplex: An Efficient SMT Solver for Verifying Deep Neural Networks},
  author    = {Katz, Guy and Barrett, Clark and Dill, David L. and Julian, Kyle and Kochenderfer, Mykel J.},
  booktitle = {Computer Aided Verification (CAV)},
  year      = {2017}
}

@article{tjeng2019mip,
  title   = {Evaluating Robustness of Neural Networks with Mixed Integer Programming},
  author  = {Tjeng, Vincent and Xiao, Kai Y. and Tedrake, Russ},
  journal = {International Conference on Learning Representations (ICLR)},
  year    = {2019}
}

@inproceedings{salman2019convex,
  title     = {A Convex Relaxation Barrier to Tight Robustness Verification of Neural Networks},
  author    = {Salman, Hadi and Yang, Greg and Zhang, Huan and Hsieh, Cho-Jui and Zhang, Pengchuan},
  booktitle = {Advances in Neural Information Processing Systems (NeurIPS)},
  year      = {2019}
}

@inproceedings{cohen2019smoothing,
  title     = {Certified Adversarial Robustness via Randomized Smoothing},
  author    = {Cohen, Jeremy M. and Rosenfeld, Elan and Kolter, J. Zico},
  booktitle = {International Conference on Machine Learning (ICML)},
  year      = {2019}
}

@inproceedings{webb2019statistical,
  title     = {A Statistical Approach to Assessing Neural Network Robustness},
  author    = {Webb, Stefan and Rainforth, Tom and Teh, Yee Whye and Kumar, M. Pawan},
  booktitle = {International Conference on Learning Representations (ICLR)},
  year      = {2019}
}

@inproceedings{yang2022ensemble,
  title     = {On the Certified Robustness for Ensemble Models and Beyond},
  author    = {Yang, Zhuolin and Li, Linyi and Xu, Xiaojun and Kailkhura, Bhavya and Xie, Tao and Li, Bo},
  booktitle = {International Conference on Learning Representations (ICLR)},
  year      = {2022}
}

@inproceedings{jia2019certified,
  title     = {Certified Robustness to Adversarial Word Substitutions},
  author    = {Jia, Robin and Raghunathan, Aditi and G{\"o}ksel, Kerem and Liang, Percy},
  booktitle = {Empirical Methods in Natural Language Processing (EMNLP)},
  year      = {2019}
}

@inproceedings{huang2019achieving,
  title     = {Achieving Verified Robustness to Symbol Substitutions via Interval Bound Propagation},
  author    = {Huang, Po-Sen and Stanforth, Robert and Welbl, Johannes and Dyer, Chris and Yogatama, Dani and Gowal, Sven and Dvijotham, Krishnamurthy and Kohli, Pushmeet},
  booktitle = {Empirical Methods in Natural Language Processing (EMNLP)},
  year      = {2019}
}

@inproceedings{ye2020safer,
  title     = {SAFER: A Structure-free Approach for Certified Robustness to Adversarial Word Substitutions},
  author    = {Ye, Mao and Gong, Chengyue and Liu, Qiang},
  booktitle = {Annual Meeting of the Association for Computational Linguistics (ACL)},
  year      = {2020}
}

@inproceedings{zeng2023certified,
  title     = {Certified Robustness to Text Adversarial Attacks by Randomized [MASK]},
  author    = {Zeng, Jiehang and Xu, Jianhan and Zheng, Xiaoqing and Huang, Xuanjing},
  booktitle = {Computational Linguistics},
  year      = {2023}
}

@inproceedings{ebrahimi2018hotflip,
  title     = {HotFlip: White-Box Adversarial Examples for Text Classification},
  author    = {Ebrahimi, Javid and Rao, Anyi and Lowd, Daniel and Dou, Dejing},
  booktitle = {Annual Meeting of the Association for Computational Linguistics (ACL)},
  year      = {2018}
}

@inproceedings{jin2020textfooler,
  title     = {Is BERT Really Robust? A Strong Baseline for Natural Language Attack on Text Classification and Entailment},
  author    = {Jin, Di and Jin, Zhijing and Zhou, Joey Tianyi and Szolovits, Peter},
  booktitle = {AAAI Conference on Artificial Intelligence},
  year      = {2020}
}

@inproceedings{morris2020textattack,
  title     = {TextAttack: A Framework for Adversarial Attacks, Data Augmentation, and Adversarial Training in NLP},
  author    = {Morris, John X. and Lifland, Eli and Yoo, Jin Yong and Grigsby, Jake and Jin, Di and Qi, Yanjun},
  booktitle = {Empirical Methods in Natural Language Processing (EMNLP), System Demonstrations},
  year      = {2020}
}

@article{zou2023gcg,
  title   = {Universal and Transferable Adversarial Attacks on Aligned Language Models},
  author  = {Zou, Andy and Wang, Zifan and Carlini, Nicholas and Nasr, Milad and Kolter, J. Zico and Fredrikson, Matt},
  journal = {arXiv preprint arXiv:2307.15043},
  year    = {2023}
}

@article{chao2023pair,
  title   = {Jailbreaking Black Box Large Language Models in Twenty Queries},
  author  = {Chao, Patrick and Robey, Alexander and Dobriban, Edgar and Hassani, Hamed and Pappas, George J. and Wong, Eric},
  journal = {arXiv preprint arXiv:2310.08419},
  year    = {2023}
}

@article{inan2023llamaguard,
  title   = {Llama Guard: LLM-based Input-Output Safeguard for Human-AI Conversations},
  author  = {Inan, Hakan and Upasani, Kartikeya and Chi, Jianfeng and Rungta, Rashi and Iyer, Krithika and Mao, Yuning and Tontchev, Michael and Hu, Qing and Fuller, Brian and Testuggine, Davide and Khabsa, Madian},
  journal = {arXiv preprint arXiv:2312.06674},
  year    = {2023}
}

@misc{detoxify,
  title  = {Detoxify},
  author = {Hanu, Laura and {Unitary team}},
  year   = {2020},
  howpublished = {\url{https://github.com/unitaryai/detoxify}}
}

@inproceedings{lin2023toxicchat,
  title     = {ToxicChat: Unveiling Hidden Challenges of Toxicity Detection in Real-World User-AI Conversation},
  author    = {Lin, Zi and Wang, Zihan and Tong, Yongqi and Wang, Yangkun and Guo, Yuxin and Wang, Yujia and Shang, Jingbo},
  booktitle = {Findings of EMNLP},
  year      = {2023}
}

@article{markov2023moderation,
  title   = {A Holistic Approach to Undesired Content Detection in the Real World},
  author  = {Markov, Todor and Zhang, Chong and Agarwal, Sandhini and Eloundou, Tyna and Lee, Teddy and Adler, Steven and Jiang, Angela and Weng, Lilian},
  journal = {AAAI Conference on Artificial Intelligence},
  year    = {2023}
}

@article{rebedea2023nemo,
  title   = {NeMo Guardrails: A Toolkit for Controllable and Safe LLM Applications with Programmable Rails},
  author  = {Rebedea, Traian and Dinu, Razvan and Sreedhar, Makesh and Parisien, Christopher and Cohen, Jonathan},
  journal = {Empirical Methods in Natural Language Processing (EMNLP), System Demonstrations},
  year    = {2023}
}

@article{kumar2023erasecheck,
  title   = {Certifying LLM Safety against Adversarial Prompting},
  author  = {Kumar, Aounon and Agarwal, Chirag and Srinivas, Suraj and Li, Aaron Jiaxun and Feizi, Soheil and Lakkaraju, Himabindu},
  journal = {arXiv preprint arXiv:2309.02705},
  year    = {2023}
}

@article{robey2023smoothllm,
  title   = {SmoothLLM: Defending Large Language Models Against Jailbreaking Attacks},
  author  = {Robey, Alexander and Wong, Eric and Hassani, Hamed and Pappas, George J.},
  journal = {arXiv preprint arXiv:2310.03684},
  year    = {2023}
}

@book{vovk2005algorithmic,
  title     = {Algorithmic Learning in a Random World},
  author    = {Vovk, Vladimir and Gammerman, Alexander and Shafer, Glenn},
  publisher = {Springer},
  year      = {2005}
}

@inproceedings{papadopoulos2002inductive,
  title     = {Inductive Confidence Machines for Regression},
  author    = {Papadopoulos, Harris and Proedrou, Kostas and Vovk, Volodya and Gammerman, Alex},
  booktitle = {European Conference on Machine Learning (ECML)},
  year      = {2002}
}

@inproceedings{papadopoulos2008normalized,
  title     = {Normalized Nonconformity Measures for Regression Conformal Prediction},
  author    = {Papadopoulos, Harris and Gammerman, Alex and Vovk, Volodya},
  booktitle = {Proceedings of the IASTED International Conference on Artificial Intelligence and Applications (AIA)},
  pages     = {64--69},
  year      = {2008}
}

@article{lei2018distribution,
  title   = {Distribution-Free Predictive Inference for Regression},
  author  = {Lei, Jing and G'Sell, Max and Rinaldo, Alessandro and Tibshirani, Ryan J. and Wasserman, Larry},
  journal = {Journal of the American Statistical Association},
  volume  = {113},
  number  = {523},
  pages   = {1094--1111},
  year    = {2018}
}

@inproceedings{vovk2012mondrian,
  title     = {Conditional Validity of Inductive Conformal Predictors},
  author    = {Vovk, Vladimir},
  booktitle = {Asian Conference on Machine Learning (ACML)},
  year      = {2012}
}

@article{barber2021limits,
  title={The Limits of Distribution-Free Conditional Predictive Inference},
  author={Rina Foygel Barber and Emmanuel J. Cand{\`e}s and Aaditya Ramdas and Ryan J. Tibshirani},
  journal={Information and Inference: A Journal of the IMA},
  year={2019},
}

@inproceedings{gendler2022rscp,
  title     = {Adversarially Robust Conformal Prediction},
  author    = {Gendler, Asaf and Weng, Tsui-Wei and Daniel, Luca and Romano, Yaniv},
  booktitle = {International Conference on Learning Representations (ICLR)},
  year      = {2022}
}

@inproceedings{devlin2019bert,
  title     = {BERT: Pre-training of Deep Bidirectional Transformers for Language Understanding},
  author    = {Devlin, Jacob and Chang, Ming-Wei and Lee, Kenton and Toutanova, Kristina},
  booktitle = {North American Chapter of the Association for Computational Linguistics (NAACL)},
  year      = {2019}
}

@article{liu2019roberta,
  title   = {RoBERTa: A Robustly Optimized BERT Pretraining Approach},
  author  = {Liu, Yinhan and Ott, Myle and Goyal, Naman and Du, Jingfei and Joshi, Mandar and Chen, Danqi and Levy, Omer and Lewis, Mike and Zettlemoyer, Luke and Stoyanov, Veselin},
  journal = {arXiv preprint arXiv:1907.11692},
  year    = {2019}
}

@article{kezins2026beyond,
  title   = {Beyond Red-Teaming: Formal Guarantees of {LLM} Guardrail Classifiers},
  author  = {Kezins, Nikita and Ekka, Urbas and Berrang, Pascal and Arnaboldi, Luca},
  journal = {arXiv preprint arXiv:2605.10901},
  year    = {2026}
}

@inproceedings{alain2017probing,
  title     = {Understanding Intermediate Layers Using Linear Classifier Probes},
  author    = {Alain, Guillaume and Bengio, Yoshua},
  booktitle = {International Conference on Learning Representations (Workshop)},
  year      = {2017}
}

@inproceedings{tenney2019bert,
  title     = {{BERT} Rediscovers the Classical {NLP} Pipeline},
  author    = {Tenney, Ian and Das, Dipanjan and Pavlick, Ellie},
  booktitle = {Annual Meeting of the Association for Computational Linguistics},
  year      = {2019}
}

@inproceedings{xin2020deebert,
  title     = {{DeeBERT}: Dynamic Early Exiting for Accelerating {BERT} Inference},
  author    = {Xin, Ji and Tang, Raphael and Lee, Jaejun and Yu, Yaoliang and Lin, Jimmy},
  booktitle = {Annual Meeting of the Association for Computational Linguistics},
  year      = {2020}
}

@inproceedings{tibshirani2019covariate,
  title     = {Conformal Prediction under Covariate Shift},
  author    = {Tibshirani, Ryan J. and Barber, Rina Foygel and Cand{\`e}s, Emmanuel J. and Ramdas, Aaditya},
  booktitle = {Advances in Neural Information Processing Systems},
  year      = {2019}
}

\newpage
\appendix
\section{Proofs}
\label{app:proofs}

Throughout, $\Delta = \phivec(x') - \phivec(X)$, and split conformal supplies
$\mathbb{P}(s(X) \le \Rad(X)) \ge 1-\alpha$ for whichever score $s$ the radius
was built from, for exchangeable calibration and test prompts. That step does not
depend on the difficulty estimate being any good, because the ratio $s/\hat{s}$
is exchangeable whatever $\hat{s}$ is; a poor $\hat{s}$ costs coverage of the
fired prompts by the certificate, never validity.

\paragraph{Proposition~\ref{prop:oracle}.}
With $\hat{s} = s$ every calibration ratio $s(X_i)/\hat{s}(X_i)$ equals $1$, so
every order statistic is $1$, $q_\alpha = 1$ and $\Rad(x) = s(x)$. (With the
floor, $\hat{s} = \max(s, 0.05)$ and the radius is slightly larger for prompts
the attacker barely moves; this is the residual between oracle CHC and the
ceiling.) A fired prompt is certified when
$h(x) > s(x) = \max_{x' \in T(x)} (h(x) - h(x'))_+$, which for $h(x) > 0$
rearranges to $\min_{x' \in T(x)} h(x') > 0$: every edit still fires. For general
$\hat{s}$, the events $E$ and $C$ split the fired prompts into four cells. Then
$1 - \mathrm{CHC} = \mathbb{P}(\bar{C}) = \mathbb{P}(\bar{E} \cap \bar{C}) +
\mathbb{P}(E \cap \bar{C})$, and $\text{ceiling} - \mathrm{CHC} = \mathbb{P}(E) -
\mathbb{P}(C) = \mathbb{P}(E \cap \bar{C}) - \mathbb{P}(\bar{E} \cap C)$. On
$\bar{E} \cap C$ some edit has $h(x') \le 0$, so $s(x) \ge h(x) > \Rad(x)$, and
the prompt is a miscoverage event. Under the oracle this cell is empty. The
oracle is not available in practice, since it uses the test prompt's own edits,
the very thing a certificate is supposed to avoid needing. \hfill$\square$

\paragraph{Proposition~\ref{prop:shared}.}
Fix a realisation on which $s_{V,M}(X) \le \Rad_{V,M}(X)$, and take any
$w \in V$ and $x' \in T(X)$. Since $w \in V$, $\Pi_V w = w$, and $\Pi_V$ is
self-adjoint, so $w \cdot \Delta = w \cdot (\Pi_V \Delta)$. Both $w$ and
$u := \Pi_V \Delta$ lie in $V$, where $M^{\pm 1/2}$ are defined and self-adjoint,
so
\begin{align*}
-w \cdot u = -(M^{1/2} w) \cdot (M^{-1/2} u)
&\le \|M^{1/2} w\|_2 \, \|M^{-1/2}\Pi_V \Delta\|_2 \\
&\le \|M^{1/2} w\|_2 \, s_{V,M}(X)
\le \|M^{1/2} w\|_2 \, \Rad_{V,M}(X) ,
\end{align*}
by Cauchy--Schwarz, the definition of $s_{V,M}$ as a maximum over $T(X)$, and
the event. Rearranging gives the claim. Since $w$ was arbitrary, the claim holds
for all of $V$ on that one event. \hfill$\square$

\paragraph{Where the displacement subspace enters.} $V$ is built inside
$\mathrm{range}(Q)$, so $\Pi_V P = 0$ and hence $\Pi_V Q = \Pi_V$: the projector
$Q$ never appears in the score. Its role is to keep $V$ away from the $k$
directions the attacker moves hardest, the limiting case of pricing them, which
the generalised eigenproblem does across the whole spectrum.

\paragraph{Proposition~\ref{prop:tight}.}
For fixed $\Delta$, maximising $-\Delta \cdot w$ over $\{w \in V :
\|M^{1/2}w\|_2 \le 1\}$ is, after the change of variable $u = M^{1/2}w$,
maximising $-(M^{-1/2}\Pi_V \Delta) \cdot u$ over the unit ball of $V$. Its value
is $\|M^{-1/2}\Pi_V\Delta\|_2$, attained at $u \propto -M^{-1/2}\Pi_V\Delta$, that
is, at $w \propto -M^{-1}\Pi_V\Delta$. The value is non-negative, so the positive
part in $s_w$ changes nothing, and the supremum over the ball equals the supremum
over its boundary. Taking the maximum over $x' \in T(x)$ of both sides and
exchanging the two maximisations, which is always allowed, gives
$s_{V,M}(x) = \sup_w s_w(x)$. A single radius valid for every unit head on one
event must be at least $s_w(x)$ for each of them, hence at least $s_{V,M}(x)$.
\hfill$\square$

\paragraph{Proposition~\ref{prop:ball}.}
Write $w = w_{\mathrm{ref}} + \delta$ with $\delta \in V$ and
$\|M^{1/2}\delta\|_2 \le \rho$. Then
$-\Delta \cdot w = -\Delta \cdot w_{\mathrm{ref}} - \Delta \cdot \delta
\le (-\Delta \cdot w_{\mathrm{ref}})_+ + \rho\|M^{-1/2}\Pi_V\Delta\|_2$,
bounding the first term by its positive part and the second by the
Cauchy--Schwarz step of Proposition~\ref{prop:shared}. The right-hand side does
not mention $w$, so its maximum over $T(X)$ is $s_\rho(X)$ for every
$w \in \mathcal{W}_\rho$, and the conformal event finishes the proof. At
$\rho = 0$ the score is $\max_{x'}(-\Delta \cdot w_{\mathrm{ref}})_+$, the per-head
score. \hfill$\square$

\paragraph{Theorem~\ref{thm:composite}.}
\emph{Validity.} $h = \max_j h_j$ is a fixed measurable function of the input,
computed by the frozen encoder, with every $\theta_j$, $\sigma_j$ and the
composite's $\hat{s}$ fixed before calibration. Split conformal applied to $s_h$
therefore gives coverage at level $\alpha$ as for any other scalar guard. Nothing
in that argument counts members, and the member attaining the maximum may depend
on $x$ freely: the selection happens inside the function, before the score is
formed.

\emph{Dominance.} Fix $x$, $x'$ and any $j$. Since $h(x') \ge h_j(x')$,
\[
h(x) - h(x') \;\le\; h(x) - h_j(x') \;=\; \big(h_j(x) - h_j(x')\big) + \big(h(x) - h_j(x)\big) ,
\]
and the second bracket is the non-negative gap $h(x) - h_j(x)$, which does not
depend on $x'$. Taking the maximum over $x' \in T(x)$ and then the minimum over
$j$ gives $s_h(x) \le \min_j [ s_j(x) + (h(x) - h_j(x)) ]$. Subtracting from
$h(x)$ gives $h(x) - s_h(x) \ge \max_j [h_j(x) - s_j(x)]$. \hfill$\square$

\paragraph{Proposition~\ref{prop:price}.}
The composite fires on a benign prompt if and only if some member does, so its
benign rate is the measure of a union of $J$ sets each of measure $\beta$. It is
at least $\beta$ and at most $\min(1, J\beta)$ by subadditivity, with equality at
the two ends when the sets coincide and when they are disjoint. The rate is
non-decreasing in $\beta$, so holding it at $a$ forces $\beta \le a$. Each $h_j$
is non-increasing in $\theta_j$, which is non-increasing in $\beta$, so every
member's margin in \eqref{eq:dominance} is evaluated at a threshold at least as
strict as a solo deployment would use. \hfill$\square$

\section{Depth: where to read, and where the geometry comes from}
\label{app:depth}

\subsection{Where the low-rank structure comes from}
\label{app:pooler}

The constructions of \S\ref{sec:shared} are worth attempting only because
displacements at $\phivec$ are low rank: $7$ of $768$ directions carry $90\%$ of
the energy an attacker moves, which leaves a quiet feature space to put heads in.
That is a property of $\phivec$. The encoder's own hidden states carry the same
displacements in $72$ to $214$ directions (Table~\ref{tab:pooler}), and one
operation sits between the last hidden state and $\phivec$.

Table~\ref{tab:pooler} attributes the compression. The pooler's learned linear
map $W_p h_\ell + b_p$ alone takes rank$_{90}$ from $74$ to $3$ at the last hidden
state, and to at most $18$ at every depth. The $\tanh$ then raises the count from
$3$ to $7$. A random projection at matched Frobenius norm
followed by the same $\tanh$ leaves $63$. The compression therefore comes from
what $W_p$ learned, which was to annihilate the directions this attacker moves.
Only $24\%$ of pooler units have $|W_p h + b_p| > 2$ at the last layer, too few for
saturation of the $\tanh$ to account for the compression.

Whether a guard admits these constructions can thus be checked before any
calibration, by comparing rank$_{90}$ of its pooled displacements with that of its
hidden-state displacements. The same observation bounds the claim: a guard with
no pooler, or with one that has not learned this invariance, should not be
expected to behave like these two.

\begin{figure}[t]
\centering
\includegraphics[width=0.46\textwidth]{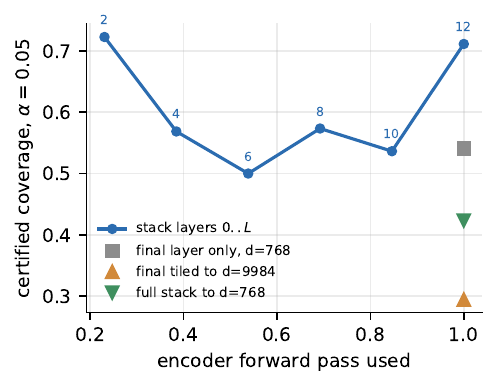}
\caption{CHC against the fraction of the encoder a guard has to run. Points on
the curve are stacks over layers $0..L$, labelled by $L$; the three markers are the
final layer alone and the two matched-dimension controls. Exiting at layer $2$
matches the all-layer guard at a quarter of the compute.}
\label{fig:budget}
\end{figure}

\subsection{Which depth to read}
\label{app:depthsweep}

The feature space of \S\ref{sec:exp:shared} was read off $\phivec$ because that is
where the shipped head sits. A frozen encoder exposes twelve more representations,
and an edit's displacement exists in all of them. The heads in this section are
trained by gradient descent with a penalty on the displacement subspace; the
certification pipeline is otherwise the one of \S\ref{sec:setup}.

\paragraph{Reading earlier certifies more.} Build a guard at each depth, with the
encoder frozen up to layer $\ell$, a linear head on the [CLS] state there, and
the same certification pipeline. CHR then peaks at layer $2$ and falls off towards
the top: $0.369$ against $0.136$ at layer $12$ under the lexical attacker, and
$0.379$ against $0.106$ under the gradient attacker. A mean-pooled control is much
flatter, peaking at $0.399$ at layer $3$ against $0.323$ at layer $12$. This is
consistent with the [CLS] position at layer $12$ being specialised for the
pretraining objective; we did not test that reading. The layer that separates
best and the layer that certifies best differ, so no accuracy result predicts
this.

\paragraph{Depth and dimension.} A single head reading every layer at once
certifies $0.711$ $[0.634, 0.782]$ against $0.541$ $[0.451, 0.631]$ for the same
head on the final layer. Under the gradient attacker the figures are $0.512$
against $0.333$, with AUROC $0.964$ against $0.731$. These intervals are marginal
and unpaired; their separation is suggestive and is not a paired test. The
all-layer head reads $9984$ features against $768$, so Table~\ref{tab:budget}
adds two controls. Tiling the final layer to $9984$ features, the same information
thirteen times over, lowers CHC to $0.295$. Tiling is not a neutral control for a
head trained by gradient descent: it multiplies the effective step size, and
above $2048$ features the displacement subspace is estimated by a randomised
sketch where the $768$-feature arm uses an exact decomposition. The drop shows
that more columns carrying the same information do not help, and it leaves the
role of capacity open. Compressing the full stack back to $768$ features with a fixed random
Gaussian gives $0.423$, so an arbitrary $768$-dimensional summary of the layers
does not carry the gain either.

\paragraph{Most of the gain costs a quarter of the compute.} A guard that exits at
layer $2$ reads $23\%$ of the encoder and certifies $0.723$ $[0.642, 0.796]$, in
line with the all-layer head's $0.711$ (Figure~\ref{fig:budget}). The curve between
them is not monotone. CHC dips to $0.500$ in the middle of the network and
recovers at the top, and the empirical ceiling tracks it at $0.854$, $0.792$ and
$0.880$, so the dip comes from robustness at those depths. AUROC moves the other way, rising from $0.902$ at layer $2$ to $0.930$ at
layer $4$ while CHC falls by $0.154$. The representation that separates best and
the one that certifies best are different.

\paragraph{Read early, intervene late.} A natural next move is to delete the
displacement subspace inside the encoder with a forward hook and let the
remaining blocks run. It does not work. Projecting at layer $\ell$ leaves a median
$0.849$ of the displacement energy present at $\phivec$ for every $\ell < 11$, and
$0.942$ when only the [CLS] position is projected; only the last block's
projection sticks, at $0.063$. The later layers rebuild the directions, and the gap
between the two scopes suggests that attention pulls part of the subspace back
from other token positions. The displacement subspaces at different depths also
differ: mean principal-angle cosines against layer $12$ run from $0.12$ to $0.62$,
and adjacent layers agree at only $0.43$ to $0.72$. Intervening inside the
encoder does not pay, which is why the constructions of \S\ref{sec:shared} live
at $\phivec$.

\begin{table}[t]
\centering
\caption{One calibration shared by every linear head in an $m$-dimensional space. Averages are over the six trained heads, none of which the calibration saw. \emph{energy in $V$} is the fraction of displacement energy the space contains, which is what the radius pays for. By Proposition~\ref{prop:tight} the shared column cannot be improved by a tighter analysis.}
\label{tab:shared}
\small
\begin{tabular}{rrrrrr}
\toprule
$m$ & energy in $V$ & AUROC & shared CHC & worst head & per-head CHC \\
\midrule
1 & 0.000 & 0.619 & 0.000 & 0.000 & 0.000 \\
2 & 0.000 & 0.618 & 0.000 & 0.000 & 0.000 \\
4 & 0.002 & 0.674 & 0.000 & 0.000 & 0.106 \\
8 & 0.002 & 0.770 & 0.000 & 0.000 & 0.120 \\
16 & 0.004 & 0.832 & 0.000 & 0.000 & 0.169 \\
32 & 0.008 & 0.849 & 0.018 & 0.010 & 0.391 \\
64 & 0.016 & 0.855 & 0.004 & 0.000 & 0.455 \\
128 & 0.028 & 0.849 & 0.000 & 0.000 & 0.463 \\
256 & 0.051 & 0.849 & 0.000 & 0.000 & 0.417 \\
512 & 0.137 & 0.859 & 0.000 & 0.000 & 0.370 \\
752 & 1.000 & 0.823 & 0.000 & 0.000 & 0.267 \\
\bottomrule
\end{tabular}
\end{table}

\begin{table}[t]
\centering
\caption{Which representation the head sits on. The last two rows are matched-dimension controls: tiling the final layer carries the information of $768$ features in $9984$, and compressing the full stack carries thirteen layers in $768$.}
\label{tab:budget}
\small
\begin{tabular}{lrrrrrr}
\toprule
representation & layers & $d$ & cost & AUROC & CHC & CHR \\
\midrule
exit at layer 2 & 3 & 2304 & 0.23 & 0.902 & \textbf{0.723} & 0.500 \\
exit at layer 4 & 5 & 3840 & 0.38 & 0.930 & 0.569 & 0.439 \\
exit at layer 6 & 7 & 5376 & 0.54 & 0.926 & 0.500 & 0.364 \\
exit at layer 8 & 9 & 6912 & 0.69 & 0.915 & 0.573 & 0.414 \\
exit at layer 10 & 11 & 8448 & 0.85 & 0.932 & 0.536 & 0.409 \\
exit at layer 12 & 13 & 9984 & 1.00 & 0.921 & 0.711 & 0.510 \\
final layer only & 1 & 768 & 1.00 & 0.906 & 0.541 & 0.333 \\
repeat final $\times 13$ & 1 & 9984 & 1.00 & 0.920 & 0.295 & 0.167 \\
stack $\to$ rp 768 & 13 & 768 & 1.00 & 0.938 & 0.423 & 0.293 \\
\bottomrule
\end{tabular}
\end{table}

\begin{table}[t]
\centering
\caption{rank$_{90}$ of the displacement matrix: how many directions carry $90\%$ of the energy an attacker moves, by depth and by representation. The last row at layer $12$ is $\varphi$ itself.}
\label{tab:pooler}
\small
\begin{tabular}{lrrrrrrr}
\toprule
representation & 2 & 4 & 6 & 8 & 10 & 11 & 12 \\
\midrule
$h_\ell$ (hidden state) & 143 & 141 & 164 & 214 & 179 & 141 & 74 \\
$W_p h_\ell + b_p$ & 18 & 6 & 2 & 2 & 5 & 3 & 3 \\
$\tanh(W_r h_\ell + b_p)$ & 118 & 119 & 136 & 166 & 146 & 117 & 63 \\
$\tanh(W_p h_\ell + b_p)$ & 24 & 20 & 22 & 19 & 22 & 13 & 7 \\
\bottomrule
\end{tabular}
\end{table}

\section{Additional measurements}
\label{app:extra}

\paragraph{Composition and measurement as substitutes.} A single depth with the
measured difficulty of \S\ref{sec:exp:shared} certifies $0.684$, within $0.016$ of
the four-layer composite, and combining the two gives $0.677$. One reading is that
both act on the slack in the difficulty estimate; \S\ref{sec:exp:which} suggests
the composite also raises margins, and these runs do not separate the two.

\paragraph{Adversarial training.} For the layer-$4$ head, adding the standard term
that requires the worst sampled edit of each training prompt to clear the margin
lowers CHC from $0.654$ to $0.512$ and the empirical ceiling from $0.797$ to
$0.715$. We report it because it is the obvious thing to try, and because the
ceiling is the quantity it was supposed to raise.

\paragraph{The shipped head.} Toxic-bert's own weight vector, restricted to the
complement of the displacement subspace and to $V$, has AUROC $0.326$ at every
$m$, below chance. Removing $16$ of $768$ directions inverts its ranking of
prompts, which suggests the shipped head relies heavily on directions the
attacker moves. This is why the six heads are re-fitted inside
$\mathrm{range}(Q)$.

\paragraph{Feature space and metric.} Replacing the generalised eigenbasis with
plain principal components costs $0.171$ of CHC ($0.455$ to $0.284$) and $0.128$
of the neighbourhood certificate at $\rho = 0.10$. It also lowers the ceiling from
$0.819$ to $0.682$, so the guard in that space is less robust as well as harder to
certify. The metric matters less: $M = I$ matches $M = \Sigma_D$ for a single head
and differs by under $0.04$ on the neighbourhood (it is better at $\rho = 0.10$),
so we report it as an ablation.

\section{Reproduction}
\label{app:repro}

Every number comes from a run artifact under \texttt{results/}. Embeddings are
produced once by \texttt{scripts/launch\_plan.py} with
\texttt{plans/latent\_shared.txt} and cached as npz files keyed on what was
embedded. The scripts below read the cache from \texttt{/var/tmp/embcache}
(\texttt{accounting.py} also honours \texttt{EMBCACHE}) and need no torch.
\begin{itemize}[leftmargin=*,itemsep=0pt,topsep=2pt]
\item \texttt{scripts/regen\_shared.py}: the certificate arms of
\S\ref{sec:exp:shared} and \S\ref{sec:exp:portability}.
\item \texttt{scripts/accounting.py}: everything in Appendix~\ref{app:review}, the
intervals and coverage in Tables~\ref{tab:difficulty} and~\ref{tab:ball}, and
Figure~\ref{fig:accounting} (left). It stops unless it reproduces every stored arm
to $10^{-9}$.
\item \texttt{scripts/head\_composite.py} (with \texttt{-{}-cache} set to the
toxic-bert lexical cache): the version composites of \S\ref{sec:exp:composite}.
\item \texttt{scripts/regen\_composite.py} (with a per-layer cache): the depth
composites.
\item \texttt{scripts/qualitative.py -{}-m 64}: \S\ref{sec:exp:which}.
\item \texttt{scripts/tex\_tables.py} and \texttt{scripts/paper\_figures.py}: the
tables and figures.
\end{itemize}
The depth sweep, the pooler attribution and the early-exit arms of
Appendix~\ref{app:depth} come from the latent stage, which needs torch.
Predictions and falsifiers were written in \texttt{PREREG\_latent.md} before the
runs they refer to existed, with four dated amendments recording which
constructions were exploratory and which predictions failed.

\paragraph{Settings.} $600$ harmful and $3000$ benign ToxicChat prompts. The
harmful prompts are split into train, calibration and test folds of $204$, $198$
and $198$; the benign prompts are used whole to fit $V$, the heads and the
thresholds. $32$ edits per prompt for the lexical attacker (up to three operations
each) and $16$ restarts for the gradient attacker, of which $8$ are spent measuring
$\hat{s}$. $k = 16$; $m = 64$; floor under $\hat{s}$ $0.05$ in raw logit units;
ridge penalty $\lambda = 1$ for the heads (one bootstrap head uses $10$) and for
the difficulty regression; shrinkage $0.01$ on $\Sigma_D$; hinge head $600$
epochs of gradient descent at step $0.05$ with penalty $10^{-3}$; benign
false-positive rate $5\%$; $\alpha = 0.05$; bootstrap intervals from $2000$
resamples. Seeds $0$ to $3$ for toxic-bert with the lexical attacker and seed $0$
for the other two settings, with \texttt{deterministic=true}.

\paragraph{Hardware and cost.} One NVIDIA H200 with an 8-core host. Embedding the
corpus dominates and happens once per guard, attacker and seed. Every result in
\S\ref{sec:exp} is then linear algebra: the twelve certificate arms regenerate in
under two minutes, and \texttt{scripts/accounting.py} runs its $34$ sweeps in
under a minute.

\section{Post-review analyses}
\label{app:review}

The analyses in this appendix were added after review, from the cached embeddings
of the runs above, and were not registered. Each arm is computed by the same code
as the stored runs, and the script stops unless it reproduces them exactly.

\paragraph{The accounting on every setting.} Table~\ref{tab:accounting} repeats
Table~\ref{tab:difficulty} across seeds, attackers and guards. The loose share
exceeds the attacker's in every row, by a factor of $1.4$ to $3.4$, and prompts
certified although an edit evades them never exceed $2.1\%$ of fired prompts in
these rows.

\paragraph{Intervals and multiplicity.} Intervals are percentile bootstraps over
test prompts with the calibration split held fixed, $2000$ resamples, and one set
of resamples per setting shared by every arm in it, so differences are paired.
They capture test-set variability. Variability across calibration splits shows up
across seeds. The three settings at seed $0$ form one family for the claim that
measuring $\hat{s}$ helps, and Table~\ref{tab:paired} adjusts its $p$-values by
Holm's method. A four-seed sign test cannot go below $p = 0.0625$, so the seeds
are reported as counts.

\paragraph{Coverage across draws.}
\label{app:coverage}
With $n = 198$ calibration prompts and $\alpha = 0.05$ the radius is the $190$th
smallest calibration ratio. Given the calibration split, miscoverage is then
$\mathrm{Beta}(9, 190)$, and the number of misses among $198$ test prompts is
beta-binomial. One split therefore delivers coverage in $[0.91, 0.99]$ with
probability $0.95$, and values below $0.95$ on one seed are expected.
Table~\ref{tab:coverage} reports coverage for every certificate on every seed. The
per-head certificates average $0.949$ to $0.956$ across seeds. The shared score on
seed $3$ is the exception: $27$ misses where $9$ are expected, an event of
probability $5.8 \times 10^{-4}$. On the same split the per-head certificates
cover at $0.934$ to $0.965$, and the neighbourhood certificate drops to $0.889$ at
$\rho = 1$, where its score is dominated by the shared term. We checked for
missing scores, for estimates at the floor and for prompts with fewer than $32$
edits, and found none, so the cause remains open.

\paragraph{Measurement budget.} Table~\ref{tab:measure} trades measurement against
the size of the certified threat model. Every row compares the measured estimate
with a ridge that certifies the same edits. The gain is $0.049$ at $K_0 = 4$, with
an interval that includes zero, and about $0.08$ at $K_0 = 8$ and $16$; the
ceiling rises with $K_0$ because fewer edits are certified.

\paragraph{The floor, its units, and the benign threshold.} Table~\ref{tab:floor}
varies the floor under $\hat{s}$. The value $0.05$ used throughout gives the
highest CHC of the three for both estimators, and the gap between them holds at
each value. The version composites of \S\ref{sec:exp:composite} apply the same
floor to $\sigma$-normalised margins, which in raw units is $0.05\sigma$ with
$\sigma \approx 0.17$; that alone moves the reference head from $0.445$ to
$0.353$. Finally, the false-positive rate is in sample (\S\ref{sec:setup}).
Fitting $V$, the heads and the thresholds on a random half of the benign prompts
gives a rate of $0.050$ on that half and $0.051$ on the other, so the threshold
transfers. CHC falls with the smaller benign set, to $0.373$ with measured
$\hat{s}$ and $0.286$ with the ridge, and the gap between them remains.

\begin{table}[t]
\centering
\caption{The accounting of Proposition~\ref{prop:oracle} on every setting and seed ($m=64$, mean over the six heads). Columns as in Table~\ref{tab:difficulty}; the last is the loose share divided by the attacker's. Prompts certified although an edit evades them (\emph{evaded}) are at most $2.1\%$ of fired prompts in every row; a two-way split would have booked them to the attacker.}
\label{tab:accounting}
\small
\begin{tabular}{llrrrrrr}
\toprule
setting & seed & $\hat{s}$ & CHC & attacker & loose & evaded & ratio \\
\midrule
toxic-bert, lexical & 0 & ridge on $\varphi$ (32 edits) & 0.374 & 0.204 & 0.422 & 0.021 & 2.1 \\
toxic-bert, lexical & 0 & ridge on $\varphi$ (24) & 0.372 & 0.161 & 0.467 & 0.020 & 2.9 \\
toxic-bert, lexical & 0 & measured (24) & 0.455 & 0.168 & 0.377 & 0.014 & 2.2 \\
\addlinespace[2pt]
toxic-bert, lexical & 1 & ridge on $\varphi$ (32 edits) & 0.290 & 0.226 & 0.484 & 0.013 & 2.1 \\
toxic-bert, lexical & 1 & ridge on $\varphi$ (24) & 0.381 & 0.216 & 0.403 & 0.015 & 1.9 \\
toxic-bert, lexical & 1 & measured (24) & 0.472 & 0.220 & 0.308 & 0.011 & 1.4 \\
\addlinespace[2pt]
toxic-bert, lexical & 2 & ridge on $\varphi$ (32 edits) & 0.342 & 0.270 & 0.388 & 0.005 & 1.4 \\
toxic-bert, lexical & 2 & ridge on $\varphi$ (24) & 0.329 & 0.244 & 0.426 & 0.003 & 1.7 \\
toxic-bert, lexical & 2 & measured (24) & 0.382 & 0.239 & 0.379 & 0.008 & 1.6 \\
\addlinespace[2pt]
toxic-bert, lexical & 3 & ridge on $\varphi$ (32 edits) & 0.312 & 0.172 & 0.516 & 0.003 & 3.0 \\
toxic-bert, lexical & 3 & ridge on $\varphi$ (24) & 0.307 & 0.157 & 0.536 & 0.002 & 3.4 \\
toxic-bert, lexical & 3 & measured (24) & 0.481 & 0.156 & 0.363 & 0.002 & 2.3 \\
\addlinespace[2pt]
toxic-bert, gradient & 0 & ridge on $\varphi$ (24) & 0.322 & 0.222 & 0.456 & 0.017 & 2.1 \\
toxic-bert, gradient & 0 & measured (24) & 0.382 & 0.236 & 0.382 & 0.003 & 1.6 \\
\addlinespace[2pt]
roberta, lexical & 0 & ridge on $\varphi$ (24) & 0.126 & 0.257 & 0.617 & 0.019 & 2.4 \\
roberta, lexical & 0 & measured (24) & 0.231 & 0.260 & 0.509 & 0.015 & 2.0 \\
\bottomrule
\end{tabular}
\end{table}

\begin{table}[t]
\centering
\caption{Measured against predicted difficulty on the same $24$ certified edits: the difference in CHC (per head, mean over six heads) and in the $\rho=0.10$ neighbourhood certificate, with $95\%$ paired bootstrap intervals over test prompts ($2000$ resamples, calibration held fixed). \emph{Holm} is the one-sided bootstrap $p$-value adjusted across the three settings at seed $0$, which form one family; seeds $1$ to $3$ are reported unadjusted. The per-head difference is positive on 4 of 4 seeds (sign test $p=0.0625$, one sided).}
\label{tab:paired}
\small
\begin{tabular}{llrrrr}
\toprule
setting & seed & $\Delta$ CHC & Holm & $\Delta$ $\rho{=}0.10$ & Holm \\
\midrule
toxic-bert, lexical & 0 & +0.083 {\scriptsize[0.019, 0.146]} & 0.016 & +0.160 {\scriptsize[0.086, 0.241]} & 0.001 \\
toxic-bert, lexical & 1 & +0.092 {\scriptsize[0.038, 0.147]} & -- & +0.217 {\scriptsize[0.117, 0.319]} & -- \\
toxic-bert, lexical & 2 & +0.052 {\scriptsize[0.001, 0.101]} & -- & +0.030 {\scriptsize[-0.047, 0.105]} & -- \\
toxic-bert, lexical & 3 & +0.174 {\scriptsize[0.115, 0.237]} & -- & +0.229 {\scriptsize[0.143, 0.317]} & -- \\
toxic-bert, gradient & 0 & +0.060 {\scriptsize[-0.001, 0.124]} & 0.027 & +0.142 {\scriptsize[0.046, 0.239]} & 0.005 \\
roberta, lexical & 0 & +0.104 {\scriptsize[0.045, 0.163]} & 0.001 & +0.020 {\scriptsize[-0.020, 0.064]} & 0.227 \\
\bottomrule
\end{tabular}
\end{table}

\begin{table}[t]
\centering
\caption{Conformal coverage $\mathbb{P}(s \le R)$ at $\alpha = 0.05$ on the 198 harmful test prompts of each seed (toxic-bert, lexical). The guarantee is marginal, over the calibration split as well as the test prompt. With $n=198$ calibration prompts, the coverage a single split delivers falls in $[0.91, 0.99]$ with probability $0.95$ (9.0 misses expected). The shared score on seed $3$ misses 27 prompts, which has probability $5.8\times 10^{-4}$ under exchangeability; the ball at large $\rho$ inherits it, since its score is then dominated by the shared term.}
\label{tab:coverage}
\small
\begin{tabular}{lrrrrr}
\toprule
certificate & seed 0 & seed 1 & seed 2 & seed 3 & mean \\
\midrule
per head, measured $\hat{s}$ & 0.943 & 0.959 & 0.971 & 0.950 & 0.956 \\
per head, ridge on $\varphi$ (24) & 0.929 & 0.940 & 0.947 & 0.980 & 0.949 \\
per head, ridge on $\varphi$ (32) & 0.933 & 0.962 & 0.938 & 0.973 & 0.951 \\
ball, $\rho=0.10$ & 0.934 & 0.970 & 0.980 & 0.934 & 0.955 \\
ball, $\rho=1.00$ & 0.944 & 0.944 & 0.965 & 0.889 & 0.936 \\
shared over $V$, $m=64$ & 0.960 & 0.949 & 0.960 & 0.864 & 0.933 \\
\bottomrule
\end{tabular}
\end{table}

\begin{table}[t]
\centering
\caption{The measurement budget. $K_0$ of the $32$ sampled edits measure $\hat{s}$ and the certificate covers the remaining $32-K_0$; the ridge baseline certifies the same edits, so each row compares like with like. The ceiling rises with $K_0$ because the threat model being certified shrinks. Toxic-bert, lexical attacker, seed $0$, $m=64$.}
\label{tab:measure}
\small
\begin{tabular}{rrrrrrrr}
\toprule
$K_0$ & certified & ceiling & measured & ridge & $\Delta$ & meas.\ $\rho{=}0.10$ & ridge $\rho{=}0.10$ \\
\midrule
0 & 32 & 0.775 & -- & 0.374 & -- & -- & 0.176 \\
4 & 28 & 0.795 & 0.411 & 0.361 & +0.049 {\scriptsize[-0.012, 0.106]} & 0.269 & 0.160 \\
8 & 24 & 0.819 & 0.455 & 0.372 & +0.083 {\scriptsize[0.019, 0.146]} & 0.328 & 0.168 \\
16 & 16 & 0.838 & 0.491 & 0.410 & +0.082 {\scriptsize[0.012, 0.144]} & 0.361 & 0.202 \\
\bottomrule
\end{tabular}
\end{table}

\begin{table}[t]
\centering
\caption{Sensitivity to the floor $\hat{s} \ge \epsilon$ that keeps the conformal ratio finite (raw logit units; the paper uses $\epsilon = 0.05$ throughout). It happens to maximise CHC for both estimators, and the gap between them holds at every value. Applied to $\sigma$-normalised margins instead, as in the version composites of \S\ref{sec:exp:composite}, the same floor is $0.05\sigma$ in raw units and the reference head certifies 0.353 instead of 0.445.}
\label{tab:floor}
\small
\begin{tabular}{rrrrrr}
\toprule
$\epsilon$ & measured & ridge & meas.\ $\rho{=}0.10$ & ridge $\rho{=}0.10$ & coverage \\
\midrule
0.01 & 0.396 & 0.326 & 0.345 & 0.185 & 0.929 \\
0.05 & 0.455 & 0.372 & 0.328 & 0.168 & 0.943 \\
0.10 & 0.419 & 0.313 & 0.294 & 0.185 & 0.940 \\
\midrule
\multicolumn{6}{l}{\footnotesize held-out benign check: fitting on 1500 benign prompts gives FPR 0.050 on them and 0.051 on the other 1500;} \\
\multicolumn{6}{l}{\footnotesize CHC with half the benign data: measured 0.373, ridge 0.286.} \\
\bottomrule
\end{tabular}
\end{table}

\end{document}